\documentclass[conference]{IEEEtran}
\IEEEoverridecommandlockouts
\usepackage{cite}
\usepackage{amsmath,amssymb,amsfonts}
\usepackage{algorithmicx}
\usepackage{graphicx}
\usepackage{textcomp}
\usepackage{xcolor}
\def\BibTeX{{\rm B\kern-.05em{\sc i\kern-.025em b}\kern-.08em
    T\kern-.1667em\lower.7ex\hbox{E}\kern-.125emX}}
    
\usepackage{subfigure}
\usepackage{tikz}
\usepackage{graphicx}
\usepackage{chemist}
\usepackage{algorithm}
\usepackage{algpseudocode}
\usepackage{empheq,etoolbox}
\usepackage{array}
\usepackage{amsmath,amsfonts,amssymb}
\usepackage{eurosym}
\usepackage{xcolor}
\usepackage{braket}
\usepackage{float}
\usepackage{svg}
\usepackage{soul}
\usepackage{hyperref}
\usepackage[app]{overpic}
\usepackage{braket}
\usepackage{bm}
\usepackage{cancel}
\usepackage{mathrsfs}
\usepackage{tikz}
\usetikzlibrary{quantikz2}
\usepackage{centernot}
\usepackage{ulem} 
\usepackage{cancel}

\usepackage{soul}

\begin{document}

\title{Refining Quantum Phase Estimation Precision Conditions on Unitaries for Many-Electron Systems}

\author{

\IEEEauthorblockN{1\textsuperscript{st}  Jérémie Messud}
\IEEEauthorblockA{
\textit{TotalEnergies}\\
2, place Jean Millier\\
92078 Paris La Défense Cedex, France\\
jeremie.messud@totalenergies.com}
\and
\IEEEauthorblockN{2\textsuperscript{nd} Wassil Sennane}
\IEEEauthorblockA{
\textit{QUANTSOC}\\
9 Rue des Colonnes\\
75002 Paris, France\\
direction@qcsocrd.eu}
}
\maketitle

\begin{abstract}
Beyond ground state energy estimation, quantum phase estimation (QPE) applied to many-electron systems has the potential to output an approximation of the ground state, enabling in a second step an evaluation of observables other than the energy. 
We here focus on the impact of approximate controlled-unitaries implementations 
on QPE precision.
After recalling the role of the QPE free parameters, we derive first‑order and unified conditions 
on the unitaries
that are necessary to control  the QPE energy estimation precision together with the QPE output state precision, important in case we want to leverage the full potential of QPE.
We apply these conditions to a Trotterization case, leading to tighter or more general bounds than in previous works. 
The main results in this article are formal. 
First numerical illustrations  on the \chemform{H_2} molecule provide useful insights.
\end{abstract}

\begin{IEEEkeywords}
Quantum phase estimation, many-electron systems, unitaries precision, Trotterization, state projection.
\end{IEEEkeywords}

\section{Introduction}

We consider the computation of bound states of many‑electron systems, governed by the stationary Schrödinger equation,
with a primary focus on the ground state.
Finding the exact solution of this equation with classical computing leads to a computational time that grows exponentially with the system size $N_S$, the latter being represented by the number of spin-orbitals when the state is expanded on an orbital basis  \cite{gao2024distributed}.
Tractable classical approaches with polynomial scaling in $N_S$, such as truncated configuration interaction (CI) \cite{zgid2012truncated}, density functional theory (DFT) \cite{kohn1965self}, and tensor‑network methods \cite{anselme2024combining,jamet2025anderson}, provide practical approximations. However, these methods can fail to achieve quantitative accuracy for strongly correlated many-electron systems, limiting their predictive capacity in computational chemistry and materials science \cite{freitas2025fundamental,lyra2022deriving,greene2022modelling}.

Quantum computing offers an alternative framework to address this challenge. In particular, quantum phase estimation (QPE) could enable the computation of ground state features
and potentially achieve exponentially superior scaling in $N_S$ than exact classical schemes \cite{shor1994algorithms,kitaev1995,nielsen2010quantum}. 
After early work established the application of QPE to electronic structure problems \cite{Abrams1999,Travaglione_2001}, various subsequent developments refined its application to molecular Hamiltonians, 
e.g. \cite{
nusran2014application,
Cruz2019OptimizingQP,
o2019quantum,
Pezz2020QuantumPE,
lin2022heisenberg,
Sugisaki2023ProjectiveMQ,
papadopoulos2024reductive,
barbieri2024multiphase,
hualde2024quantum,
shukla2025practicalquantumphaseestimation
}.
In addition to the estimation of ground state energy (denoted by ground energy in the following to lighten the notations), QPE has the potential to output a ground state approximation \cite{Travaglione_2001} (called QPE `projection' capability), enabling in a second step an evaluation of observables other than the energy.

The practical performance of QPE critically depends  on several free parameters. These include the number of phase qubits $N$, 
the number of required shots, the overlap of the initial state $\ket{\psi_{\rm init}}$ with the exact ground state, and the implementation of controlled-unitaries that involve an exponentiation of the system Hamiltonian $H$ together with a free parameter $t$. In practice, these latter implementation often involves a Trotterization approximation, introducing an additional free parameter called the number of Trotter steps $n$ which provides control on the trade‑off between precision (or error) and circuit depth \cite{suzuki_general_1991,rajagopal_generalization_1999,CHEHADE2026430, ronfaut2026numerical}. 
{A deep understanding of these parameters and their impact 
is important} for an efficient use of QPE in computational chemistry and material science.
{Reference  \cite{sennane2026robustnessquantumphaseestimation} presents a review of the QPE free parameters with  an emphasis on aspects related to the success probability.}
Complementarily, precision aspects related to an approximate implementation of unitaries remain to be explored further. 
Indeed, {existing precision} conditions are known to be quite loose \cite{childs_theory_2021}, which can lead to an overestimation of the required QPE resources, and specifically related to energy estimation \cite{mehendale_estimating_2025}.

After recalling the QPE free parameters that are significant for our considerations,
we build on the works of  \cite{childs_theory_2021,mehendale_estimating_2025} and detail a first-order formalism as a unified basis to develop precision conditions on unitaries.
We then propose new conditions that contribute to control energy estimation precision together with output state precision, important to leverage the projection potential of QPE.
We finally apply these conditions to a Trotterization case and show that they lead to bounds that are tighter than those obtained in \cite{childs_theory_2021} and more general than those obtained in \cite{mehendale_estimating_2025}.
Our main results are formal. A first numerical test on the \chemform{H_2} molecule provides useful insights.

\section{QPE overview, conditions and problem}\label{sec:method}

We consider the stationary Schrödinger equation:
\begin{equation}
H \ket{\psi_j} = E_j \ket{\psi_j},
\end{equation}
indexed by a positive integer $j$.
$E_0$ and $\ket{\psi_0}$ denote the ground energy and ground state, respectively.

QPE uses two registers \cite{shor1994algorithms, nielsen2010quantum}.  
The first one is an $N$-qubit phase register $\ket{l}$ whose measure encodes an estimation of an eigen-energy of the system.
The second one is an $N_S$-qubit system register $\ket{\psi_{\rm init}}$,
related to a spin-orbital expansion.
It can be expressed using the basis of eigenstates of $H$ as:
\begin{equation}\label{eq:init}
\ket{\psi_{\rm init}} = \sum_{j\ge 0} c_j \ket{\psi_j}
\quad,\quad
\sum_{j\ge 0} |c_j|^2=1.
\end{equation}
In practice, $\ket{\psi_{\rm init}}$ is obtained from a prior computation with polynomial cost, e.g. truncated CI  \cite{zgid2012truncated}, tensor networks \cite{anselme2024combining, jamet2025anderson}, or parameterized quantum circuits \cite{fomichev2024initial,halder2021digitalquantumsimulationstrong}.
We define the following unitary operator that acts on the system register:
\begin{equation}\label{U}
  U = e^{-i2\pi t H},
\end{equation}
where the usage of Hartree (Ha) units for energy and atomic units for time $t$ is implicit.
QPE leverages a series of unitaries $U^{2^q}$ each controlled by the phase qubit number $q \in \{0,...,N-1\}$.
A  phase register measure that gives a value $l\in\{0,...,2^N-1\}$ projects the system register on the following output state:
\begin{equation}\label{ampj}
\ket{\psi^{(l)}_{out}} =
\sum_{j\ge 0} c_j^{(l)} \ket{\psi_j}
\quad,\quad
c_j^{(l)} = c_j \frac{f\left(\theta_j-\frac{l}{2^N}\right)}{\sqrt{P(l)}},
\end{equation}
where the probability associated to the measure is given by:
\begin{equation}\label{Prob}
P(l)=\sum_{j\ge 0} \left|c_j\right|^2 \left|f\left(\theta_j-\frac{l}{2^N}\right) \right|^2, 
\end{equation}
and $
    f\left(\theta_j-\frac{l}{2^N}\right)
    = \frac{1}{2^{N}} \frac{\sin\left(\pi 2^N (\theta_j-\frac{l}{2^N})\right)}{\sin\left(\pi (\theta_j-\frac{l}{2^N})\right)}e^{i \pi (2^N - 1) (\theta_j-\frac{l}{2^{N}})}
$
represents a `blurring function' related to discretization effects,
due to the fact that QPE approximates a continuous phase $\theta_j\in[0,1[$ by a discrete quantity $\frac{l}{2^N}$
\cite{Travaglione_2001, bauer2025postvariationalgroundstateestimation,sennane2026robustnessquantumphaseestimation}.

QPE can allow us to approximate $E_0$ from the most probable  phase  measurements $l^*$ in the sense of $P(l)$, equation (\ref{Prob}), provided an initial state with a sufficiently large $|c_0|^2$ is given and discretization effects are mitigated \cite{Travaglione_2001,sennane2026robustnessquantumphaseestimation}.
We suppose that we are in such a situation in the following.
The ground energy is then estimated by ($\lceil \cdot\rceil$ is the ceiling function):
\begin{align}\label{E_j}
E_0 &\approx   -\frac{1}{t}\frac{l^*}{2^N} + \frac{1}{t}\lceil E_0 t \rceil,
\end{align}
which requires an a priori knowledge of $\lceil E_0 t \rceil$ (this will be discussed later).
After a $l^*$  measurement, the output state $\ket{\psi^{(l^*)}_{out}}$ can provide a projection on the ground state $\ket{\psi_0}$ 
in the situations discussed in \cite{Travaglione_2001,sennane2026robustnessquantumphaseestimation}, i.e. when $|c_0^{(l^*)}|^2\gg|c_{j>0}^{(l^*)}|^2$.

QPE's efficiency 
relies on various free parameters \cite{sennane2026robustnessquantumphaseestimation}. 
We remind conditions on the parameters that {are important for the} developments here, related to unitaries implementation.

    \textbf{($t$ condition.)} 
    The time $t$ choice is the first parameter to set as it is crucial to allow to recover the ground energy $E_0$ from $\frac{l^*}{2^N}$, i.e. to constrain \textit{a priori} the $\lceil E_0 t \rceil$ value in (\ref{E_j}).
    Given that the problem Hamiltonian is naturally expressed as a Linear Combination of Unitaries (LCU) \cite{nielsen2005fermionic, childs2012hamiltonian, loaiza2023reducing},
        \begin{align}\label{eq:LCU}
        H=\sum_{\beta=1}^M \gamma_{\beta} H_\beta
        \quad,\quad \gamma_\beta\in \mathbb{R},
        \end{align}
        where $H_\beta$ are Hermitian unitaries,
        it is common to choose
            $t=\frac{\alpha}{\sum_\beta |\gamma_\beta|}$ where
            $\alpha\in\left[\frac{1}{2},1\right]$, which implies
            $\lceil E_0 t \rceil=0$.
        Two other cases leveraging prior information and  leading to larger $t$ values were derived in  \cite{sennane2026robustnessquantumphaseestimation}.
    %
    
    \textbf{($N$ condition.)} Once $t$ defined and the ambiguity on $\lceil E_0 t \rceil$ removed, we can set accordingly the number of phase qubits. Reaching { chemical} precision $\varepsilon_{\rm ch}=1.6\times10^{-3} \text{ Ha}$ on energy estimation requires:
    \begin{align}\label{Nt}
    N \geq N_{min}(t)=\left\lceil\log_2{\left(\frac{1}{t\varepsilon_{\rm ch}}\right)}\right\rceil -1.
    \end{align}
    Interestingly, the minimum number of phase qubits $N_{\rm min}(t)$ is directly related to the choice made for $t$, a smaller $t$ leading to a larger $N_{\rm min}(t)$. 
    Adding more phase qubits, i.e. taking
    $N = N_{\rm min}(t) +a$ where $a$ is a positive integer,
    can make sense to mitigate discretization effects or increase the confidence level, as detailed in  \cite{sennane2026robustnessquantumphaseestimation}.
    
    \textbf{($U^{2^{q}}$ condition.)} The practical and approximate implementation of $U^{2^{q}}$ is denoted by $\mathcal{S}(U^{2^q})$.
    The following condition {constrains achieving} a precision driven by $\varepsilon$ on the approximate unitaries ($\lVert \cdot \rVert_{2}$ denotes the spectral norm):
    \begin{equation}\label{U_app}
        \lVert U^{2^q} - \mathcal{S}(U^{2^q}) \rVert_{2} \lesssim 2\pi t2^q \varepsilon \quad=\quad \frac{2^q}{2^{N_\text{min}(t)}}\pi{\text{ if $\varepsilon=\varepsilon_{\rm ch}$}},
    \end{equation}
    the equality being obtained using (\ref{Nt}). {As we will demonstrate, choosing $\varepsilon=\varepsilon_{\rm ch}$ can be related to first-order to the ability of the approximate unitaries to achieve chemical precision on energy, but smaller $\varepsilon$ choices can make sense to also control output state precision.}
    In the case of order-$p$ Trotterization for $\mathcal{S}(U^{2^q})$, an upper bound on the minimum number $n_\text{min}^{(p)}(q,t)$ of Trotter steps to reach precision $\varepsilon$ in (\ref{U_app}) is given by \cite{sennane2026robustnessquantumphaseestimation,childs_theory_2021,mehendale_estimating_2025}:
    \begin{equation}\label{nmin}
        {n_\text{min}^{(p)}(q,t) 
        \approx
              \left\lceil 2\pi 2^{q}t \left(C_p /\varepsilon\right)^\frac{1}{p}\right\rceil},
    \end{equation}
    where forms for $ C_p $  will be given later.
    Then, if $\varepsilon=\varepsilon_{\rm ch}$ is constrained,
    we have $n_\text{min}^{(p)}(q,t) \approx \lceil \frac{2^q}{2^{N_\text{min}(t)}} \pi(\frac{ C_p }{\varepsilon^{p+1}})^\frac{1}{p}\rceil$,
    and a minimum number of Trotter steps over the whole QPE circuit,
    \begin{equation}\label{nmin-max}
    \begin{gathered}
    n_\text{min-tot}^{(p)}\approx\left\lceil {2^a}\mathscr{C}_p \right\rceil,
    \quad \mathscr{C}_p =
    \pi\left( C_p /\varepsilon^{p+1}\right)^\frac{1}{p},
    \end{gathered}
    \end{equation}
    that tends to be independent of $N_\text{min}(t)$ or $t$, and to depend only on the number $a$ of phase qubits beyond $N_\text{min}(t)$. The corresponding QPE complexity equals $\left\lceil {2^a}\mathscr{C}_p \right\rceil\times \mathscr{N}_p(N_S)$, where $\mathscr{N}_p(N_S)$ represents one Trotter step complexity.

However, challenges {related to the precision of approximate implementations of unitaries} remain.
The work in  \cite{childs_theory_2021} leads to a specific $C_p$ form 
which overestimates $n_\text{min}^{(p)}$ and thus the required QPE resources \cite{mehendale_estimating_2025}. The work in \cite{mehendale_estimating_2025} leads to a tighter bound but it is energy specific. To our knowledge, {conditions to control} the output state precision remain to be studied, which is important in case we want to leverage the projection potential of QPE.
We here build on the works of \cite{childs_theory_2021,mehendale_estimating_2025} and detail a unified first-order formalism that allows us to 
propose new conditions that contribute to control QPE energy estimation and output state precisions together.

\section{Unitary operator precision condition}
\label{sec:trott}


The `approximate' $\mathcal{S}(U^{2^q})$ must be unitary to be implementable in a quantum framework.
We consider it is defined through an `effective QPE Hamiltonian' $H^\mathcal{S}(\lambda)$ (Hermitian):
\begin{equation}\label{eq: def_H_s_lambda}
\mathcal{S}(U^{2^q})=e^{-i 2\pi t 2^q H^\mathcal{S}(\lambda)}\quad,\quad\lambda\in\mathbb{R}^+,
\end{equation}
where the $\lambda$ parameter controls the `quality' of the effective Hamiltonian, $H^\mathcal{S}(\lambda)$ getting closer to $H$ as $\lambda$ gets smaller. 
In other terms, $H^\mathcal{S}(\lambda)$ is defined by the Hermitian logarithm of the approximate unitary (a Trotterization case will be detailed below).
We denote by $E_{i}^{\mathcal{S}}(\lambda)$ the eigen-energies of $H^\mathcal{S}(\lambda)$
and by  $\ket{ \psi_{i}^\mathcal{S}(\lambda) }$ its eigen-states, which represent the `effective QPE energies and states' (resp.).
We do not explicit the $\lambda$ dependency of $\mathcal{S}(U^{2^q})$ to keep the notations light.

As $H$ and $H_\mathcal{S}(\lambda)$ are Hermitian, corollary A.5 of   \cite{childs_theory_2021} gives:
\begin{align}\label{eq:unit}
    \lVert U^{2^q} - \mathcal{S}(U^{2^q})\rVert_{2} 
    \le  2\pi t 2^{q} \lVert H-H^\mathcal{S}(\lambda)\rVert_{2},
\end{align}
which is not directly interpretable as a precision condition on energy $E_{i}^{\mathcal{S}}(\lambda)$ or state $\ket{ \psi_{i}^\mathcal{S}(\lambda) }$.
To obtain such an interpretation,  we consider in the following a first-order development of $H^\mathcal{S}(\lambda)$ in $\lambda$, accurate for sufficiently small  $\lambda$ values (we will come back to the pertinency of this assumption later):
\begin{equation}
H^\mathcal{S}(\lambda) = H + \lambda \delta H + \mathcal{O}\left(\lambda^2\right)
,\quad
\delta H = \delta H^\dagger.
\label{eq: Heff_pert_series}
\end{equation}
From  (\ref{eq:unit})-(\ref{eq: Heff_pert_series}), we have:
\begin{align}
    &\lVert U^{2^q} - \mathcal{S}(U^{2^q})\rVert_{2} 
    \le  2\pi t 2^{q} \lVert \lambda \delta H\rVert_{2} + \mathcal{O}\left(\lambda^2\right),
\end{align}
which leads to the following first-order condition:
\begin{align}
\label{eq:cond_unit}
\boxed{
\begin{aligned}
    & \frac{1}{2\pi t 2^{q}}\lVert U^{2^q} - \mathcal{S}(U^{2^q})\rVert_{2} 
    \lesssim \lVert \lambda \delta H\lVert_{2}\\ 
    &\text{(unitary condition to $1^{st}$-order in $\lambda$)},
\end{aligned}
}
\end{align}
that upper-bounds the unitaries difference.
All $\approx$ (instead of $=$) and $\lesssim$ (instead of $\le$) in this article highlight that a first-order approximation has been considered to obtain the result.

\section{Energy precision condition} 
\label{sec:energy}

We aim at controlling the eigen-energy precision related to the effective QPE Hamiltonian, important for an usage of QPE for energy estimation.
From now on, we consider perturbation theory for a non-degenerate state $i$.
Note that the other states $j\ne i$ can be degenerate without affecting any of our results (we thus do not make potential degeneracy explicit, which lightens the notations). 
In the end, we will focus on $i=0$ 
meaning that we will suppose a non-degenerate ground state,
which is preferable in the context of an usage of QPE for projection \cite{sennane2026robustnessquantumphaseestimation}.

Perturbation theory gives
$E_{i}^{\mathcal{S}}(\lambda) = E_{i}+ \bra{ \psi_{i} } \lambda \delta H \ket{\psi_{i}} + \mathcal{O}(\lambda^2)$
(chapter XI of \cite{cohen2018mecanique}), which leads to:
\begin{equation}
\label{eq:Ei_firts}
\lvert E_{i}^{\mathcal{S}}(\lambda) - E_{i} \rvert
= \lvert \bra{\psi_{i}} \lambda \delta H \ket{\psi_i} \rvert +  \mathcal{O}(\lambda^2).
\end{equation}

To bound the energy difference, we establish the relation:
\begin{align}\label{eq:varineq}
\lvert \bra{\psi_{i}}  \delta H \ket{\psi_{i}} \rvert \le 
\sqrt{ \bra{\psi_i} ( \delta H)^2 \ket{\psi_i} }
\le 
\lVert \delta H \rVert_{2},
\end{align}
where the first inequality is obvious and the second inequality is related to the definition of the spectral norm for a Hermitian operator, i.e. $\lVert \delta H \rVert_{2}=\max_{\lVert \ket{\psi} \rVert_{2} =1} \sqrt{ \bra{\psi} ( \delta H)^2 \ket{\psi} }$.

From  (\ref{eq:Ei_firts}) and (\ref{eq:varineq}), we deduce:
\begin{align}
\label{eq:cond_unit3}
\boxed{
\begin{aligned}
    &|E_i^{\mathcal S}(\lambda) - E_i| 
    \lesssim \sqrt{ \bra{\psi_i} (\lambda \delta H)^2 \ket{\psi_i} }\le \|\lambda\,\delta H\|_{2} \\
    &\quad\quad\text{ (energy condition to $1^{\mathrm{st}}$-order in $\lambda$)}.
\end{aligned}
}
\end{align}
This gives options to upper-bound the energy difference.

\section{State precision condition} 

We aim at controlling the precision of the eigen-state  $\ket{ \psi_{i}^\mathcal{S}(\lambda) }$  related to the effective QPE Hamiltonian, necessary in the context of a usage of QPE for projection.
A reasoning directly on states is impractical due to their dimensionality. Scalar measures of distinguishability between states
like the fidelity or trace distance \cite{nielsen2010quantum} are commonly used.

Following \cite{GU_2010, Rossini_2018},  perturbation theory leads to~
%
\footnote{
Standard perturbation theory leads to states $\ket{ \psi_{i}^{\mathcal{S}}(\lambda) } =  \ket{ \psi_{i} } -  \lambda \sum_{k \ne i}\frac{ \bra{\psi_{k}} \delta H \ket{\psi_{i}} }{E_k-E_i} \ket{ \psi_{k} } + \mathcal{O}(\lambda^2)$ that are normalized to first-order in $\lambda$,
i.e. $\langle\psi_{i}^{\mathcal{S}}(\lambda)|\psi_{i}^{\mathcal{S}}(\lambda)\rangle=1+\mathcal{O}(\lambda^2)$ and $\langle\psi_{i}|\psi_{i}^{\mathcal{S}}(\lambda)\rangle=1$. This is not suitable for our problematic. The solution is to consider strict normalization perturbation theory \cite{GU_2010,Rossini_2018}. This affects state expansion but not energy expansion, i.e. all results derived in section \ref{sec:energy} remain the same, which is coherent.
}:
\begin{align}
&&\lvert \braket{\psi_{i} | \psi_{i}^{\mathcal{S}}(\lambda)} \rvert
=
1-\frac{\lambda^2}{2} \sum_{k\ne i}\frac{ \lvert \bra{\psi_{k}} \delta H \ket{\psi_{i}} \rvert^2}{(E_k-E_i)^2}+\mathcal{O}(\lambda^4)\nonumber\\
&&\Rightarrow
\lvert \braket{\psi_{i} | \psi_{i}^{\mathcal{S}}(\lambda)} \rvert^2
=
1-\lambda^2 \sum_{k\ne i}\frac{\lvert \bra{\psi_{k}} \delta H \ket{\psi_{i}} \rvert^2}{(E_k-E_i)^2}+\mathcal{O}(\lambda^4).\nonumber
\end{align}
This allows us to evaluate among others the 
trace distance, equal in our case to the square root of one minus the fidelity:
\begin{align}\label{eq:T2}
T(\ket{ \psi_{i} }, \ket{ \psi_{i}^{\mathcal{S}}(\lambda)} ) 
&=
\sqrt{1-\lvert \braket{\psi_{i} | \psi_{i}^{\mathcal{S}}(\lambda)} \rvert^2} \\
&= \lambda\sqrt{\sum_{k\ne i}\frac{\lvert \bra{\psi_{k}} \delta H \ket{\psi_{i}} \rvert^2}{(E_k-E_i)^2}}+\mathcal{O}(\lambda^2).\nonumber
\end{align}
%
Note that, contrariwise to the energy, 
having $\lambda$ small is not sufficient to ensure a small first-order term for trace distance.
We must additionally have:
$\forall k \ne i: \lvert E_k-E_i \rvert \centernot\ll \lambda \lvert \bra{\psi_{k}} \delta H \ket{\psi_{i}} \rvert$, making state precision conditions usually more stringent than energy conditions.

To bound the trace distance, we establish the relation:
\begin{align}\label{eq:demoA}
    \sqrt{\sum_{k\ne i}\frac{\lvert \bra{\psi_{k}} \delta H \ket{\psi_{i}} \rvert^2}{(E_k-E_i)^2}}
    \le
    \frac{1}{\Delta E_i}\sqrt{\sum_{k\ne i} \lvert \bra{\psi_{k}} \delta H \ket{\psi_{i}} \rvert^2}\\
    =
    \frac{1}{\Delta E_i}\sqrt{ \bra{\psi_{i}} (\delta H)^2 \ket{\psi_{i}} - \bra{\psi_{i}} \delta H \ket{\psi_{i}}^2}
    \nonumber\\
    =
    \frac{A_i}{\Delta E_i}\sqrt{ \bra{\psi_{i}} (\delta H)^2 \ket{\psi_{i}}}\le 
    \frac{A_i}{\Delta E_i}\lVert \delta H \rVert_{2},
    \nonumber
\end{align}
where we considered the $i^{th}$-state energy gap for the first inequality:
\begin{align}\label{eq: varepsilon}
\Delta E_i =\underset{k\ne i}{\min}\rvert E_k - E_i \rvert,
\end{align}
the reasoning in chapter XI of  \cite{cohen2018mecanique} for the standard-deviation equality, the following quantity for the second equality (defined for $\langle\psi_{i}| (\delta H)^2|\psi_{i}\rangle\ne 0$ which is not constraining
\footnote{
Having $\bra{\psi_i} (\delta H)^2 \ket{\psi_i} = \lVert \delta H \ket{\psi_i} \rVert_{2}^2=0$
implies $\delta H \ket{\psi_i}=0$, thus $\forall n\ge 1: (\delta H)^n \ket{\psi_i}=0$, leading to perturbation theory irrelevance for state $i$. 
}):
\begin{align}\label{eq:A0}
   A_i  =  \sqrt{1-\frac{\bra{\psi_{i}} \delta H \ket{\psi_{i}}^2}{\bra{\psi_{i}} (\delta H)^2 \ket{\psi_{i}}}}\in[0,1],
\end{align}
and the relation in \eqref{eq:varineq} for the last inequality.

From  (\ref{eq:T2}) and (\ref{eq:demoA}), we deduce:
\begin{align}
\label{eq:dHproj}
\boxed{
\begin{aligned}
    & \frac{\Delta E_i}{A_i} \times T(|\psi_{i}\rangle,|\psi_{i}^{\mathcal{S}}(\lambda)\rangle) \\
    &\hspace{2.5cm}\lesssim \sqrt{\bra{\psi_{i}} (\lambda \delta H)^2 \ket{\psi_{i}}}
    \le \lVert \lambda \delta H\lVert_{2}\\
    &\quad\quad\quad\text{ }\text{ (state condition to $1^{st}$-order in $\lambda$)}.
\end{aligned}
}
\end{align}
This gives options to upper-bound the state precision.

\section{Constraining energy precision and state precision together}

From  (\ref{eq:cond_unit}), (\ref{eq:cond_unit3}) and (\ref{eq:dHproj}), we can deduce that constraining:
\begin{align}
\label{eq:cond_unit_unif1}
\boxed{\begin{aligned}
     \|\lambda\,\delta H\|_{2}&\le \varepsilon\quad
     \text{(unitary \& energy \& state)},
\end{aligned}
}
\end{align}
allows us to upper-bound in one go the precisions on unitaries, energy difference and trace distance (for any state $i$).

Then, we can deduce using (\ref{eq:cond_unit3}) and (\ref{eq:dHproj}) that both chemical precision $\varepsilon_{\rm ch}$ on energy difference and target precision $\alpha_\text{tar}$ on trace distance (for state $i$) can be controlled taking:
\begin{align}\label{eq:varepsilon2}
\boxed{
\begin{aligned}
    \quad\varepsilon=\min\left(\varepsilon_{\rm ch}, \frac{\Delta E_i}{A_i}\alpha_\text{tar}\right)
    \quad\text{(energy \& state)}.
\end{aligned}
}
\end{align}
%
%
Note that doing the choice $\varepsilon=\varepsilon_{\rm ch}$
instead of \eqref{eq:varepsilon2} 
is equivalent to constrain the target precision of state $i$ to a specific  precision value $\alpha_{{\rm ch}_i}$ given by:
\begin{align}\label{eq:alpha.ch}
   \alpha_{\rm tar}\quad\rightarrow \quad\alpha_{{\rm ch}_i}=\frac{A_i}{\Delta E_i}\varepsilon_{\rm ch}.
\end{align}
In the context of an usage of QPE for projection in addition to energy estimation, this choice yields a reasonable $\alpha_{{\rm ch}_i}$ when
$\frac{A_i}{\Delta E_i}\le 1$ Ha$^{-1}$, since then $\alpha_{{\rm ch}_i}\lesssim 10^{-3}$,
corresponding to a good trace-distance precision. By contrast, if
$\frac{A_i}{\Delta E_i}\ge 10$ Ha$^{-1}$, then $\alpha_{{\rm ch}_i}>10^{-2}$, leading to a much poorer precision. In such cases, the usage of the condition in  (\ref{eq:varepsilon2}) is required to give control on both energy and trace distance precisions.

We highlight that the upper-bounds related to \eqref{eq:cond_unit_unif1} can be loose. 
Interestingly,  we could define from  (\ref{eq:cond_unit3}), (\ref{eq:dHproj}) and \eqref{eq:varepsilon2} a tighter upper-bound that can still allow us to control in one go the precisions on energy difference and trace distance:
\begin{align}
\label{eq:cond_unit_unif2}
\boxed{
\begin{aligned}
     \sqrt{ \bra{\psi_i} ( \lambda \delta H)^2 \ket{\psi_i} }&\le \varepsilon
     \hspace{3mm}\text{(energy \& state, tighter)},
\end{aligned}
}
\end{align}
but not anymore the precision on unitaries difference as $\sqrt{ \bra{\psi_i} ( \lambda\delta H)^2 \ket{\psi_i} }$ does not appear in (\ref{eq:cond_unit}).
Equation \eqref{eq:cond_unit_unif2} leads to a tighter bound than \eqref{eq:cond_unit_unif1}
when the eigen-state $i$ is far from the `maximum state' that defines the spectral norm.

The bound related to (\ref{eq:cond_unit_unif2}) can still be loose. Indeed, the trace distance condition demonstration considered $\Delta E_i$ instead of the larger $E_{k\ne i}-E_i$ values, remind (\ref{eq:demoA}), which can hardly be changed for practical reasons.
But, for the energy difference, (\ref{eq:Ei_firts}) may lead us to consider (if the $1^{st}$-order term is not zero):
\begin{align}
\label{eq:cond_unit_unif3}
\boxed{
\begin{aligned}
     \lvert \bra{\psi_{i}} \lambda \delta H \ket{\psi_i} \rvert\lesssim \varepsilon
     \hspace{3mm}\text{(energy only, even tighter)},
\end{aligned}
}
\end{align}
which does not anymore constrain a precision on unitaries and trace distance,
as $\lvert \bra{\psi_{i}} \lambda \delta H \ket{\psi_i} \rvert$ cannot be directly related to these two quantities.
This is why it is sufficient to consider only the chemical precision $\varepsilon=\varepsilon_{\rm ch}$ in (\ref{eq:cond_unit_unif3}), instead of the more constraining form (\ref{eq:varepsilon2}).
Note that \cite{mehendale_estimating_2025} proposed a condition similar to (\ref{eq:cond_unit_unif3}) in the context of order-$2$ Trotterization.

All the conditions developped above are valid for any state $i$. From now on we consider the ground state $i=0$, which is usually the main target of QPE applied many-electron systems. $\Delta E_0$, equation (\ref{eq: varepsilon}), then represents the spectral gap.
It is to be noted from (\ref{eq:varepsilon2}) that adding the state precision condition can usually be more constraining than using the energy precision condition only,  since the spectral gap $\Delta E_0$ can often be small for realistic many-electron systems.
Equation (\ref{eq:varepsilon2}) raises the question of estimating \textit{a priori} a reasonable value for $\frac{\Delta E_0}{A_0}$ (e.g., from the system giving the QPE initial state or statistical analysis).
Similarly, the potential usage of conditions  (\ref{eq:cond_unit_unif2}) or (\ref{eq:cond_unit_unif3}) raises the question of estimating a reasonable approximation of $\ket{\psi_0}$ \cite{mehendale_estimating_2025}.
Note that the spectral norm condition (\ref{eq:cond_unit_unif1}) can be efficiently approximated numerically using stochastic estimation when $\delta H$ is defined by a LCU \cite{blunt_monte_2025}, 
which might represent an advantage.
An extensive discussion of these aspects goes beyond the scope of this article.

This achieves to detail a unified formalism that allowed us to justify and develop various first-order precision conditions related to approximate implementation of unitaries.
Especially, the conditions (\ref{eq:dHproj})-(\ref{eq:varepsilon2}) and (\ref{eq:cond_unit_unif2}) are new to our knowledge.
We now explicit clear forms for $\lambda \delta H$ on Trotterization cases.

\section{Trotterization cases}

Given that $H$ naturally takes a LCU form ~\eqref{eq:LCU}, we have:
\begin{equation}
    U^{2^q} = 
    e^{-i2\pi t 2^{q}H} = 
    e^{-i2\pi t 2^q\sum_{\beta=1}^M \gamma_\beta H_{\beta}},
\end{equation}
which encourages the usage of order-$p$ Trotterization approximation, denoted by $\mathcal{S}(U^{2^q}, p)$, a common implementation 
where a parameter $n\in\mathbb{N}^*$ called the `number of Trotter steps' must be set \cite{nielsen2010quantum,suzuki_general_1991, rajagopal_generalization_1999}.
From the Baker-Campbell-Hausdorff (BCH) formula \cite{nielsen2010quantum}, we have \cite{childs_theory_2021,mehendale_estimating_2025}:
\begin{align}\label{lambd_p}
&\mathcal{S}(U^{2^q},p)=e^{-i2\pi t 2^q H^\mathcal{S}_p(\lambda_p) }
,\quad
\lambda_p=\left(2\pi \frac{t}{n} 2^q\right)^p,
\end{align}
where $H^\mathcal{S}_p(\lambda_p)=H+\lambda_p \delta H_p +\mathcal{O}(\lambda_p^2)$ represents the effective order-$p$ Trotter Hamiltonian (not to be confused with the order of our previous developments in  $\lambda$).
Equations \eqref{eq:cond_unit_unif1}, \eqref{eq:cond_unit_unif2} or \eqref{eq:cond_unit_unif3} can each be rewritten in the form $\lambda_p C_p\lesssim\varepsilon$, where:
\begin{align}
\label{eq:C_p}
\boxed{
\begin{aligned}   
    C_p \,\  & \rightarrow \,\ \lVert  \delta H_p\lVert_{2} 
   \,\ \text{ or }
   \sqrt{ \bra{\psi_0} ( \delta H_p)^2 \ket{\psi_0} }\text{ (smaller)}\\
   &\text{ or }
   \lvert \bra{\psi_0} \delta H_p \ket{\psi_0} \rvert\text{ (energy only, even smaller)}
   .
\end{aligned}
}
\end{align}
Thus,  after some manipulations using (\ref{lambd_p}), we can deduce that ensuring 
the precision conditions \eqref{eq:cond_unit_unif1}, \eqref{eq:cond_unit_unif2} or \eqref{eq:cond_unit_unif3} implies:
\begin{align}
\label{eq:n-min}
    &n\gtrsim n_\text{min}^{(p)}(q,t),
\end{align}
where $n_\text{min}^{(p)}(q,t)$ is defined by (\ref{nmin}) and $C_p$ can take the
explicit forms in (\ref{eq:C_p}) in the cases that have been justified here.
In other terms, using at least $n_\text{min}^{(p)}(q,t)$ Trotter steps computed from one of the  $C_p$  in (\ref{eq:C_p}) ensures that the conditions in  (\ref{eq:cond_unit}), (\ref{eq:cond_unit3}) or (\ref{eq:dHproj}) are satisfied, depending on the chosen option
\footnote{
It is possible in \eqref{Nt} to consider the $\varepsilon$ of \eqref{eq:varepsilon2}  instead of $\epsilon_\text{ch}$. This constrains $N_\text{min}(t)$ to be larger than necessary to reach chemical precision on energy estimation, but the qubit overhead remains small (log scaling). The conceptual interest is that the property in (\ref{nmin-max}) then always holds, i.e. a minimum number of Trotter steps over the whole circuit that tends to depend only on $a$.
}.

Note that the work in  \cite{childs_theory_2021} can be considered as giving a different and always larger $C_p$ value, that we denote by $C'_p$, obtained from considerations directly on $\lVert U^{2^q} - \mathcal{S}(U^{2^q}, p)\rVert_{2}$
without involving first-order approximation in $\lambda_p$.
Our first-order reasoning leads to tighter upper-bounds, which might theoretically have a more limited range of validity 
but tend to behave better in practice 
(a numerical illustration will be given in next section).
Let us make that explicit using the $\delta H_p$  form given by order-$1$ Trotter approximation \cite{childs_theory_2021,mehendale_estimating_2025}:
\begin{align}\label{trotter_1st_init}
&\mathcal{S}(U^{2^q}, 1)^\frac{1}{n}= \prod_{\beta=1}^M e^{-i H_\beta \gamma_\beta\frac{2\pi t 2^q}{n}},\quad \lambda_1=2\pi \frac{t}{n} 2^q \\
&\delta H_1 = -\frac{i}{2} \sum_{\alpha=1}^{M-1} \sum_{\beta=\alpha+1}^M \left[ \gamma_\beta H_\beta, \gamma_\alpha H_\alpha \right].\nonumber
\end{align}
Whatever choice made in (\ref{eq:C_p}), we have:
\begin{align}\label{trotter_1st}
C_1 \,\ \le \,\ \lVert  \delta H_1\lVert_{2} = \frac{1}{2}\Big\lVert \text{ }\sum_{\alpha=1}^{M-1} \sum_{\beta=\alpha+1}^M \left[ \gamma_\beta H_\beta, \gamma_\alpha H_\alpha \right] \text{ } \Big\rVert_{2},
\end{align}
whereas the upper-bound obtained in \cite{childs_theory_2021} leads to :
\begin{align}\label{trotter_1st-bis}
C_1' = \frac{1}{2}\sum_{\alpha=1}^{M-1}\Big\lVert \text{ } \sum_{\beta=\alpha+1}^M \left[ \gamma_\beta H_\beta, \gamma_\alpha H_\alpha \right] \text{ }\Big\rVert_{2}.
\end{align}
By triangular inequality, we always have $C_1\le C'_1$.
A similar conclusion holds for any order-p Trotterization, i.e. $C_p\le C'_p$ where $C'_p$ is the form obtained in \cite{childs_theory_2021}
\footnote{
Order-$2$ Trotter approximation implies $\lambda_2=\left(2\pi \frac{t}{n} 2^q\right)^2$ and $\delta H_2=-\frac{1}{3}\sum_{\alpha=1}^{2M-1}\sum_{\beta=\alpha+1}^{2M}\sum_{\nu=\beta}^{2M}\left(1-\frac{\delta_{\nu,\beta}}{2}\right)[\gamma_{\nu}H_{\nu},[\gamma_{\beta}H_{\beta},\gamma_{\alpha}H_{\alpha}]]$ with $H_{M+i} = H_{M+1-i}$ for $i = 1, ..., M$.
We can demonstrate that the $C'_2$ formulation in  \cite{childs_theory_2021} differs from $||\delta H_2||_2$ by the `position' of  $\sum_{\alpha=1}^{2M-1}$ (outside the norm in $C'_2$), leading by triangular inequality to $C_2\le C'_2$, etc.
}.
We now illustrate numerically various upper-bounds discussed in this article.

\section{Order-$1$ Trotterization and \chemform{H_2} molecule\label{sec:4}}

We consider the \chemform{H_2} molecule in the configuration summarized in Table \ref{tab:intial_data_H2}. All calculations were performed on a Quantum Learning Machine (QLM) from Bull, which enables manipulation of molecular Hamiltonians as well as emulations of quantum processing units using the myQLM package.
Exact eigen-energies $E_j$ and eigen-states $\ket{\psi_j}$ were obtained by a full diagonalization of the Hamiltonian. This allowed us to calculate $\Delta E_0$, $A_0$, etc. Spectral norms are computed with the SciPy package \cite{2020SciPy-NMeth}.
Order-$1$ Trotterization ($p=1$) is used, as in \eqref{trotter_1st_init}.
QPE is initialized with the Hartree-Fock (HF) state, $\ket{\psi_{\rm init}}=\ket{\psi_{\rm HF}}$ (with energy $E_{\rm HF}$), good enough to ensure QPE's success w.r.t. energy estimation and state projection.

\begin{table}[!t]
    \centering
    \caption{Data for \chemform{H_2} in Ha and atomic units with bond length $0.5\,\text{\AA}$. The STO-3G basis set is used: each \chemform{H} is represented with a 1s orbital, leading to a system with $N_S=4$ qubits.}
    \label{tab:intial_data_H2}
    \begin{tabular}{l l r}
        \hline
        &\textbf{Parameter} & \textbf{Value} \\
        \hline
        Initial system &$E_{\rm init}$ & $-1.042996$ \\
        Exact system &$E_0$ & $-1.055160$ \\
        &$\Delta E_0$ & $0.702985$ \\
        QPE parameters& $t=1/(2\sum_\beta \lvert\gamma_\beta\rvert)$ & $0.215149$ \\
        &$\lceil E_0\, t\rceil=\lceil E_{\rm init}\, t \rceil$ & $0$ \\
        &$N_{\rm min}(t)$ & $11$ \\
        Order-$1$ Trott. features &$A_0$ & ${1}$ \\ 
        &$||\delta H_1||_2$=$\sqrt{\lvert \bra{\psi_0} ( \delta H_1)^2 \ket{\psi_0} \rvert}$ & ${0.052420}$ \\
        &$C_1^\prime$ & $0.196930$ \\
        &$n_{\min}^{(1)}(0,t)$ & $30$ \\
        &$n_{\text{min-tot}}^{(1)}(a=0)$ & $6\times 10^{4}$ \\
        \hline
    \end{tabular}
\end{table}

\begin{figure}[h!]
    \centering
    \includegraphics[scale=0.41]{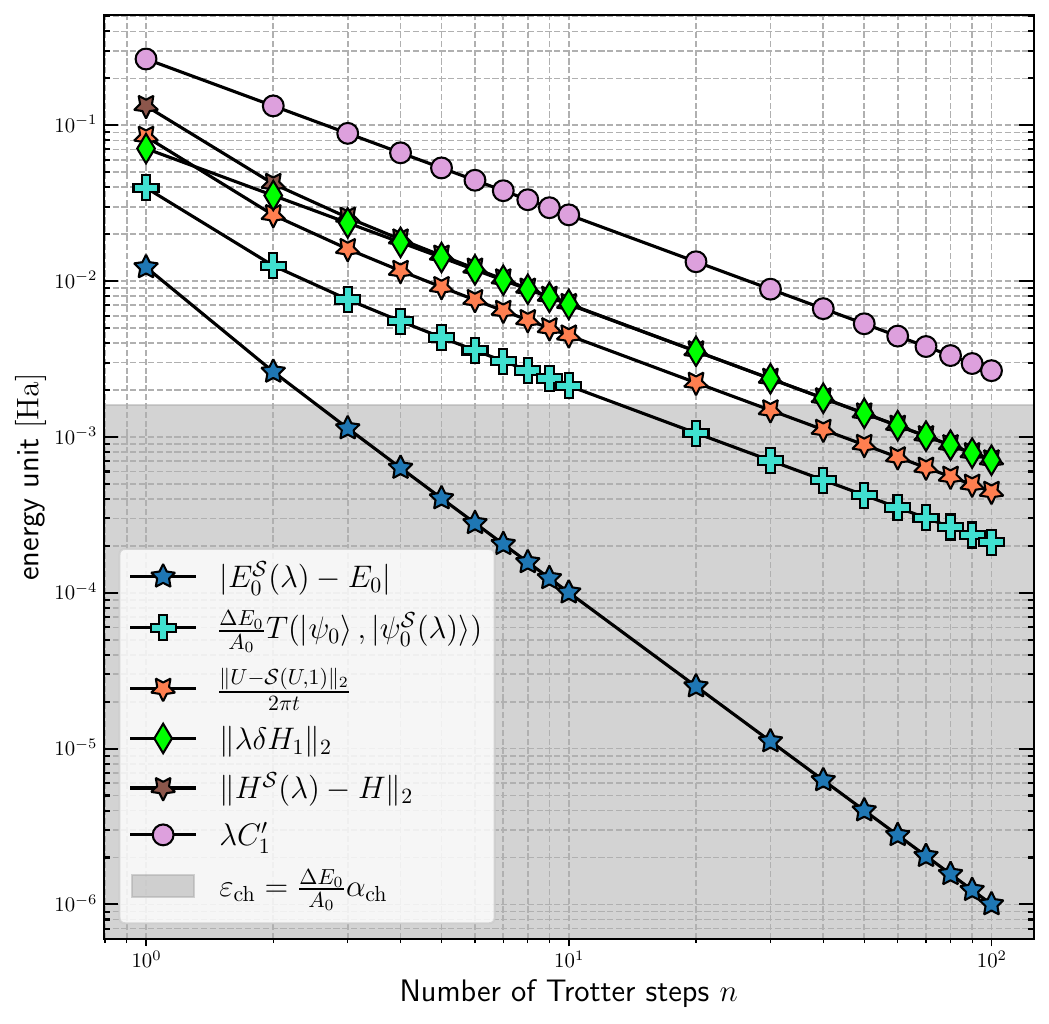}
    \caption{QPE results for \chemform{H_2} ground state with $t=1/(2\sum_\beta \lvert\gamma_\beta\rvert)$. Plot of the quantities related to \eqref{eq:unit}, \eqref{eq:cond_unit}, \eqref{eq:cond_unit3}, \eqref{eq:dHproj}, \eqref{eq:C_p} and \eqref{trotter_1st-bis} (see caption), as a function of the Trotter steps number $n$. 
    Trace distance and unitaries difference are rescaled on an energy unit, with factors $\Delta E_0/A_0$ and $1/(2\pi t)$ (resp.).
    The chemical precision area appears in grey; it can be related to trace distance precision  $\alpha_{{\rm ch}_0}\le 2.3\times10^{-3}$ (unitless) through  \eqref{eq:alpha.ch}.
    }
    \label{fig:fig_Trotterization_H2_E_spectral_as_ntr_all}
\end{figure}
\begin{figure*}[!]
    \centering
    \includegraphics[scale=0.54]{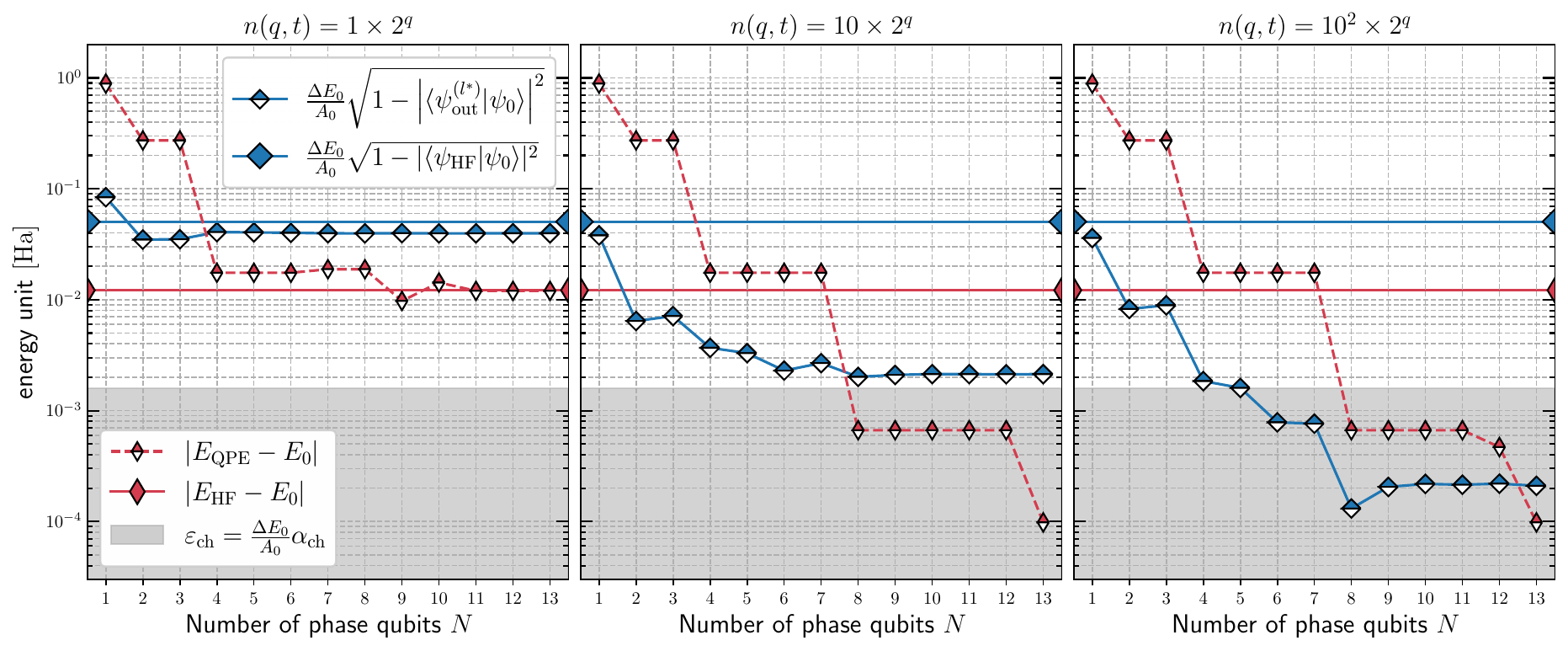}
    \put(-460, 192){(a)}
    \put(-308, 192){(b)}
    \put(-155, 192){(c)}
    \caption{QPE results for \chemform{H_2} ground state with $t=1/(2\sum_\beta \lvert\gamma_\beta\rvert)$. Plots \textbf{(a), (b), (c)} are function of the phase qubits  number  $N$ and correspond to various  Trotter step numbers  $n(q,t)$ ($n$ should be considered as a function of $q$ and $t$ 
    in coherence with (\ref{eq:n-min})).
    The evolution of energy differences (red) and trace distance (blue, rescaled on an energy using $\Delta E_0/A_0$) is shown. Straight lines correspond to initial (HF) ground state quantities.
    The chemical precision area appears in grey; it can be related to trace distance precision  $\alpha_{{\rm ch}_0}\le 2.3\times10^{-3}$ (unitless)  through  \eqref{eq:alpha.ch}.
    }
    \label{fig_QPE_H2_as_ntr}
\end{figure*}

Fig.~\ref{fig:fig_Trotterization_H2_E_spectral_as_ntr_all} illustrates the quantities derived above, all rescaled on an energy unit (Ha) to ease comparison, for $q=0$ (thus $\lambda=2\pi \frac{t}{n}$). 
It can first be seen that $\lvert E_0^\mathcal{S}(\lambda) - E_0\rvert$ does not evolve in the same way than other quantities. Indeed, 
we have in this specific \chemform{H_2} case $\bra{\psi_0}  \delta H_1 \ket{\psi_0} = 0$ which implies $A_0 = 1$ and a cancellation of the first-order energy term in (\ref{eq:Ei_firts}), that invalidates the use of the option (\ref{eq:cond_unit_unif3}) 
(note that \cite{mehendale_estimating_2025} used (\ref{eq:cond_unit_unif3})  for \chemform{H_2} in the context of order-$2$ Trotterization, which is valid as we then have $|\bra{\psi_0}  \delta H_2 \ket{\psi_0}|\ne 0$).
This leads for the energy difference to a second-order behavior in $\lambda= 2\pi \frac{t}{n}$, i.e. $\propto (\frac{1}{n})^2$, as observed. 
All other quantities evolve as a first-order in $\lambda$ thus in $\frac{1}{n}$. 
The curves associated to $\lVert H^\mathcal{S}(\lambda)-H\rVert_{2}$ and it's first-order approximation $\lVert  \delta H_1\lVert_{2}$
show that sticking with $\lVert  \delta H_1\lVert_{2}$ is largely sufficient as soon as $n \geq 5$. Adding higher-order terms would be needed for small values of $n$ (thus large values of $\lambda = 2\pi \frac{t}{n}$) but these values are not pertinent as they are far from allowing us to reach the chemical precision. The rescaled unitary difference $\lVert U - \mathcal{S}(U,1)\rVert_{2} $ is of the same order of magnitude than $\lVert  \delta H_1\lVert_{2}$, which is satisfying.
We notice two significant gaps in the curves. The first  gap is related to the difference between $C_1^\prime$, equation~\eqref{trotter_1st-bis}, and $C_1\rightarrow \lVert  \delta H_1\lVert_{2}$, equation~\eqref{trotter_1st}, which reduces by almost an order of magnitude the required number of Trotter steps $n$ to reach a desired precision and can be useful for more accurate resource estimation. Indeed, $C_1^\prime \approx 4 \lVert  \delta H_1\lVert_{2}$, which illustrates that the `triangular inequality effect' is non-negligible, especially since this effect is likely to increase with the system size  $N_S$ (implying in general more LCU terms).
Note that the usually tighter bound in \eqref{eq:cond_unit_unif2}  satisfies in our \chemform{H_2} case the uncommon relation $\sqrt{ \bra{\psi_0} ( \delta H_1)^2 \ket{\psi_0} }=\lVert  \delta H_1\lVert_{2}$
(illustrations on molecules for which the equality does not hold will be presented in future publications).
The second gap lies in the difference between the rescaled trace distance and the energy term, leading to more than an order of magnitude difference on the required $n$ to reach a similar precision.
The main reason, already explained above, is that the energy difference behaves as a second-order. More generally, trace distance may tend to converge more slowly than energy difference because of spectral gap $\Delta E_0$ effects.

Fig. \ref{fig_QPE_H2_as_ntr} illustrates the QPE behavior for various numbers of phase qubits and Trotter steps. We can observe how these parameters allow us to improve the initial state features and reach a desired precision, coherently with our previous considerations.
However, the difference between the QPE estimated ground energy and the exact ground energy does not evolve as strongly as in Fig. \ref{fig:fig_Trotterization_H2_E_spectral_as_ntr_all} w.r.t. Trotter step numbers.
This is mostly due to our limited $N$ range, which gives a lower-bound on the attainable energy precision, and thus limits the  energy difference evolution between panels \textbf{(b)} and \textbf{(c)}. 
If we added more phase qubits, the energy difference would continue to decrease coherently with Fig.~\ref{fig:fig_Trotterization_H2_E_spectral_as_ntr_all}. 
On the other hand, the trace distance between the QPE projected state (after a $l^*$ measure) and the exact ground state
seems to reach the Trotterization limit of Fig. \ref{fig:fig_Trotterization_H2_E_spectral_as_ntr_all}. 
Indeed, if the $N$ choice affects directly the energy estimation, it affects less directly the projected state.
This is because projection is driven by the $c_j \times f(\theta_j-\frac{l^*}{2^N})/\sqrt{P(l^*)}$ terms in \eqref{ampj}, which are mostly impacted by $N$ changes through 
the discretization effects
depicted in \cite{Travaglione_2001,sennane2026robustnessquantumphaseestimation}.
The slightly erratic behavior of the trace distance w.r.t. $N$ is related to these effects (they remain limited as the initial state is here good).

\section{Conclusion}

Overall, this work contributes to providing a clearer picture of the limitations and opportunities of QPE applied to many-electron systems.
We focused on a study of the precision  related to approximate implementations of the unitaries. 
We developed various first-order conditions that contribute to control the ground energy estimation precision together with the output state precision, important to leverage the full potential of QPE.
The role of the spectral gap $\Delta E_0$ and of $A_0$ on state precision tends to make  requirements on unitaries more stringent for  projection than for energy estimation. 
Applied to Trotterization, our first‑order upper bounds are tighter than those obtained in \cite{childs_theory_2021} or more general than those obtained in \cite{mehendale_estimating_2025}.
Our first numerical test on \chemform{H_2} illustrated these behaviors.

Numerical studies on various many-electron systems are being carried out, among others to estimate the statistical variations of the $A_0/\Delta E_0$ term 
and evaluate the impact of the options  \eqref{eq:cond_unit_unif1}, \eqref{eq:cond_unit_unif2} and \eqref{eq:cond_unit_unif3}.
Numerical methods to  efficiently approximate the involved quantities
are being evaluated.

\section*{Acknowledgment}
We warmly thank Y. Chatterjee, H. Calandra, J.-P. Mascom\`ere and O. Ladhari for insightful comments. We thank TotalEnergies for the permission to publish this work.









\bibliography{bibli}

@article{ronfaut2026numerical,
      title={Numerical Error Extraction by Quantum Measurement Algorithm}, 
      author={Clement Ronfaut and Robin Ollive and Stephane Louise},
      year={2026},
      journal={arXiv: 2602.01927},
      url={https://arxiv.org/abs/2602.01927}, 
}

@ARTICLE{2020SciPy-NMeth,
  author  = {Virtanen, Pauli and Gommers, Ralf and Oliphant, Travis E. and
            Haberland, Matt and Reddy, Tyler and Cournapeau, David and
            Burovski, Evgeni and Peterson, Pearu and Weckesser, Warren and
            Bright, Jonathan and {van der Walt}, St{\'e}fan J. and
            Brett, Matthew and Wilson, Joshua and Millman, K. Jarrod and
            et al and {SciPy 1.0 Contributors}},
  title   = {{{SciPy} 1.0: Fundamental Algorithms for Scientific
            Computing in Python}},
  journal = {Nature Methods},
  year    = {2020},
  volume  = {17},
  pages   = {261--272},
  adsurl  = {https://rdcu.be/b08Wh},
  doi     = {10.1038/s41592-019-0686-2},
}

@book{cohen2018mecanique,
  title={M{\'e}canique Quantique - Tome 2: Nouvelle {\'e}dition},
  author={Cohen-Tannoudji, C. and Diu, B. and Laloe, F.},
  isbn={9782759822850},
  url={https://books.google.fr/books?id=9KxxDwAAQBAJ},
  year={2018},
  publisher={EDP sciences}
}

@article{CHEHADE2026430,
title = {Error estimates and higher order Trotter product formulas in Jordan-Banach algebras},
journal = {Linear Algebra and its Applications},
volume = {730},
pages = {430-449},
year = {2026},
issn = {0024-3795},
doi = {https://doi.org/10.1016/j.laa.2025.10.032},
url = {https://www.sciencedirect.com/science/article/pii/S0024379525004458},
author = {Sarah Chehade and Andrea Delgado and Shuzhou Wang and Zhenhua Wang}
}

@article{sennane2026robustnessquantumphaseestimation,
      title={On the robustness of Quantum Phase Estimation to compute ground properties of many-electron systems}, 
      author={Wassil Sennane and Jérémie Messud},
      year={2026},
      journal={arXiv:2601.05788, submitted for publication},
      url={https://arxiv.org/abs/2601.05788}
}

@article{GU_2010,
   title={FIDELITY APPROACH TO QUANTUM PHASE TRANSITIONS},
   volume={24},
   ISSN={1793-6578},
   url={http://dx.doi.org/10.1142/S0217979210056335},
   DOI={10.1142/s0217979210056335},
   number={23},
   journal={International Journal of Modern Physics B},
   publisher={World Scientific Pub Co Pte Ltd},
   author={GU, SHI-JIAN},
   year={2010},
   month=sep, pages={4371–4458} }

@article{Rossini_2018,
   title={Ground-state fidelity at first-order quantum transitions},
   volume={98},
   ISSN={2470-0053},
   url={http://dx.doi.org/10.1103/PhysRevE.98.062137},
   DOI={10.1103/physreve.98.062137},
   number={6},
   journal={Physical Review E},
   publisher={American Physical Society (APS)},
   author={Rossini, Davide and Vicari, Ettore},
   year={2018},
   month=dec }

@article{Abrams1999,
  title = {Quantum Algorithm Providing Exponential Speed Increase for Finding Eigenvalues and Eigenvectors},
  author = {Abrams, Daniel S. and Lloyd, Seth},
  journal = {Phys. Rev. Lett.},
  volume = {83},
  issue = {24},
  pages = {5162--5165},
  numpages = {0},
  year = {1999},
  month = {Dec},
  publisher = {American Physical Society},
  doi = {10.1103/PhysRevLett.83.5162},
  url = {https://link.aps.org/doi/10.1103/PhysRevLett.83.5162}
}

@misc{kitaev1995,
      title={Quantum measurements and the Abelian Stabilizer Problem}, 
      author={A. Yu. Kitaev},
      year={1995},
      eprint={quant-ph/9511026},
      archivePrefix={arXiv},
      primaryClass={quant-ph},
      url={https://arxiv.org/abs/quant-ph/9511026}, 
}

@article{zgid2012truncated,
  title={Truncated configuration interaction expansions as solvers for correlated quantum impurity models and dynamical mean-field theory},
  author={Zgid, Dominika and Gull, Emanuel and Chan, Garnet Kin-Lic},
  journal={Physical Review B—Condensed Matter and Materials Physics},
  volume={86},
  number={16},
  pages={165128},
  year={2012},
  publisher={APS},
  doi={https://doi.org/10.1103/PhysRevB.86.165128},
}

@article{hualde2024quantum,
  title={Quantum computing in corrosion modeling: Bridging research and industry},
  author={Hualde, Juan Manuel Aguiar and Kowalik, Marek and Remme, Lian and Wolff, Franziska Elisabeth and van Velzen, Julian and Killick, Walden and Bottcher, Rene and Weimer, Christian and Krauser, Jasper and Marsili, Emanuele},
  journal={arXiv preprint arXiv:2412.07933},
  year={2024},
  doi={https://doi.org/10.48550/arXiv.2412.07933},
  url={https://arxiv.org/abs/2412.07933}, 
}

@article{Cruz2019OptimizingQP,
  title={Optimizing quantum phase estimation for the simulation of Hamiltonian eigenstates},
  author={P. M. Q. Cruz and Gonçalo Catarina and Ronan Gautier and Joaqu'in Fern'andez-Rossier},
  journal={Quantum Science \& Technology},
  year={2019},
  volume={5},
  url={https://api.semanticscholar.org/CorpusID:204512213}
}

@article{Pezz2020QuantumPE,
  title={Quantum Phase Estimation Algorithm with Gaussian Spin States},
  author={Luca Pezz{\`e} and Augusto Smerzi},
  journal={PRX Quantum},
  year={2020},
  url={https://api.semanticscholar.org/CorpusID:222208816}
}

@article{Sugisaki2023ProjectiveMQ,
  title={Projective Measurement-Based Quantum Phase Difference Estimation Algorithm for the Direct Computation of Eigenenergy Differences on a Quantum Computer},
  author={Kenji Sugisaki},
  journal={Journal of Chemical Theory and Computation},
  year={2023},
  volume={19},
  pages={7617 - 7625},
  url={https://api.semanticscholar.org/CorpusID:264437588}
}

@article{papadopoulos2024reductive,
    author = "Papadopoulos, Nicholas J. C. and Reilly, Jarrod T. and Wilson, John Drew and Holland, Murray J.",
    title = "{Reductive quantum phase estimation}",
    eprint = "2402.04471",
    archivePrefix = "arXiv",
    primaryClass = "quant-ph",
    doi = "10.1103/PhysRevResearch.6.033051",
    journal = "Phys. Rev. Res.",
    volume = "6",
    number = "3",
    pages = "033051",
    year = "2024"
}

@article{barbieri2024multiphase,
   title={Quantum multiphase estimation},
   volume={65},
   ISSN={1366-5812},
   url={http://dx.doi.org/10.1080/00107514.2025.2469974},
   DOI={10.1080/00107514.2025.2469974},
   number={2},
   journal={Contemporary Physics},
   publisher={Informa UK Limited},
   author={Barbieri, Marco and Gianani, Ilaria and Goldberg, Aaron Z. and Sánchez-Soto, Luis L.},
   year={2024},
   month=apr, pages={112–124} }

@phdthesis{nusran2014application,
  title={Application of phase estimation algorithms to improve diamond spin magnetometry},
  author={Nusran, Naufer},
  year={2014},
  school={University of Pittsburgh}
}

@article{o2019quantum,
  title={Quantum phase estimation of multiple eigenvalues for small-scale (noisy) experiments},
  author={O’Brien, Thomas E and Tarasinski, Brian and Terhal, Barbara M},
  journal={New Journal of Physics},
  volume={21},
  number={2},
  pages={023022},
  year={2019},
  publisher={IOP Publishing},
  doi = {10.1088/1367-2630/aafb8e},
  url = {https://doi.org/10.1088/1367-2630/aafb8e},
}

@article{lin2022heisenberg,
  title={Heisenberg-limited ground-state energy estimation for early fault-tolerant quantum computers},
  author={Lin, Lin and Tong, Yu},
  journal={PRX quantum},
  volume={3},
  number={1},
  pages={010318},
  year={2022},
  publisher={APS},
  doi = {10.1103/PRXQuantum.3.010318},
  url = {https://link.aps.org/doi/10.1103/PRXQuantum.3.010318},
}

@article{kohn1965self,
  title={Self-consistent equations including exchange and correlation effects},
  author={Kohn, Walter and Sham, Lu Jeu},
  journal={Physical review},
  volume={140},
  number={4A},
  pages={A1133},
  year={1965},
  publisher={APS},
  doi={https://doi.org/10.1103/PhysRev.140.A1133},
}

@article{jamet2025anderson,
  title={Anderson impurity solver integrating tensor network methods with quantum computing},
  author={Jamet, Fran{\c{c}}ois and Lindoy, Lachlan P and Rath, Yannic and Lenihan, Connor and Agarwal, Abhishek and Fontana, Enrico and Simkovic, Fedor and Martin, Baptiste Anselme and Rungger, Ivan},
  journal={APL Quantum},
  volume={2},
  number={1},
  year={2025},
  publisher={AIP Publishing},
  doi={https://doi.org/10.1063/5.0245488},
}

@article{anselme2024combining,
  title={Combining matrix product states and noisy quantum computers for quantum simulation},
  author={Anselme Martin, Baptiste and Ayral, Thomas and Jamet, Fran{\c{c}}ois and Ran{\v{c}}i{\'c}, Marko J and Simon, Pascal},
  journal={Physical Review A},
  volume={109},
  number={6},
  pages={062437},
  year={2024},
  publisher={APS},
  doi={https://doi.org/10.1103/PhysRevA.109.062437},
}

@article{gao2024distributed,
  title={Distributed implementation of full configuration interaction for one trillion determinants},
  author={Gao, Hong and Imamura, Satoshi and Kasagi, Akihiko and Yoshida, Eiji},
  journal={Journal of Chemical Theory and Computation},
  volume={20},
  number={3},
  pages={1185--1192},
  year={2024},
  publisher={ACS Publications},
  doi = {10.1021/acs.jctc.3c01190},
  url = {https://doi.org/10.1021/acs.jctc.3c01190},
}

@book{nielsen2010quantum,
  title={Quantum computation and quantum information},
  author={Nielsen, Michael A and Chuang, Isaac L},
  year={2010},
  publisher={Cambridge university press}, 
  doi={https://doi.org/10.1017/CBO9780511976667},
}

@inproceedings{shor1994algorithms,
  title={Algorithms for quantum computation: discrete logarithms and factoring},
  author={Shor, Peter W},
  booktitle={Proceedings 35th annual symposium on foundations of computer science},
  pages={124--134},
  year={1994},
  organization={Ieee},
  url = {https://doi.org/10.1109/SFCS.1994.365700},
  doi = {10.1109/SFCS.1994.365700},
}

@article{nielsen2005fermionic,
  title={The Fermionic canonical commutation relations and the Jordan-Wigner transform},
  author={Nielsen, Michael A and others},
  journal={School of Physical Sciences The University of Queensland},
  volume={59},
  pages={75},
  year={2005}, 
  url={https://api.semanticscholar.org/CorpusID:199373281},
}

@article{childs2012hamiltonian,
  title={Hamiltonian simulation using linear combinations of unitary operations},
  author={Childs, Andrew M and Wiebe, Nathan},
  journal={arXiv:1202.5822},
  year={2012}, 
  doi={https://doi.org/10.26421/QIC12.11-12},
}

@article{freitas2025fundamental,
  title={Fundamental of CO2 Adsorption and Diffusion in Subnanoporous Materials: Application to CALF-20},
  author={Freitas Goncalves, Andre de and Parazzi Lyra, Emerson and Ramdas Chavan, Sayali and Llewellyn, Philip L and Mercier Franco, Luis Fernando and Magnin, Yann},
  journal={The Journal of Physical Chemistry C},
  volume={129},
  number={40},
  pages={18190--18199},
  year={2025},
  publisher={ACS Publications}, 
  doi={https://doi.org/10.1021/acs.jpcc.5c03130},
}

@article{lyra2022deriving,
  title={Deriving force fields with a multiscale approach: From ab initio calculations to molecular-based equations of state},
  author={Lyra, Emerson P and Franco, Lu{\'\i}s FM},
  journal={The Journal of Chemical Physics},
  volume={157},
  number={11},
  year={2022},
  publisher={AIP Publishing},
  doi={https://doi.org/10.1063/5.0109350},
}

@article{greene2022modelling,
  title={Modelling carbon capture on metal-organic frameworks with quantum computing},
  author={Greene-Diniz, Gabriel and Manrique, David Zsolt and Sennane, Wassil and Magnin, Yann and Shishenina, Elvira and Cordier, Philippe and Llewellyn, Philip and Krompiec, Michal and Ran{\v{c}}i{\'c}, Marko J and Ramo, David Mu{\~n}oz},
  journal={EPJ Quantum Technology},
  volume={9},
  number={1},
  pages={37},
  year={2022},
  publisher={Springer Berlin Heidelberg}, 
  doi={https://doi.org/10.1140/epjqt/s40507-022-00155-w},
}

@misc{bauer2025postvariationalgroundstateestimation,
      title={Post-Variational Ground State Estimation via QPE-Based Quantum Imaginary Time Evolution}, 
      author={Nora Bauer and George Siopsis},
      year={2025},
      eprint={2504.11549},
      archivePrefix={arXiv},
      primaryClass={quant-ph},
      url={https://arxiv.org/abs/2504.11549}, 
}

@article{Travaglione_2001,
   title={Generation of eigenstates using the phase-estimation algorithm},
   volume={63},
   ISSN={1094-1622},
   url={http://dx.doi.org/10.1103/PhysRevA.63.032301},
   DOI={10.1103/physreva.63.032301},
   number={3},
   journal={Physical Review A},
   publisher={American Physical Society (APS)},
   author={Travaglione, B. C. and Milburn, G. J.},
   year={2001},
   month=feb }

@misc{halder2021digitalquantumsimulationstrong,
      title={Digital quantum simulation of strong correlation effects with iterative quantum phase estimation over the variational quantum eigensolver algorithm: $\mathrm{H_4}$ on a circle as a case study}, 
      author={Dipanjali Halder and Srinivasa Prasannaa V. and Valay Agarawal and Rahul Maitra},
      year={2021},
      eprint={2110.02864},
      archivePrefix={arXiv},
      primaryClass={quant-ph},
      url={https://arxiv.org/abs/2110.02864}, 
}

@article{fomichev2024initial,
  title={Initial state preparation for quantum chemistry on quantum computers},
  author={Fomichev, Stepan and Hejazi, Kasra and Zini, Modjtaba Shokrian and Kiser, Matthew and Fraxanet, Joana and Casares, Pablo Antonio Moreno and Delgado, Alain and Huh, Joonsuk and Voigt, Arne-Christian and Mueller, Jonathan E and others},
  journal={PRX Quantum},
  volume={5},
  number={4},
  pages={040339},
  year={2024},
  publisher={APS}, 
  doi={https://doi.org/10.1103/PRXQuantum.5.040339},
}

@article{loaiza2023reducing,
  title={Reducing molecular electronic Hamiltonian simulation cost for linear combination of unitaries approaches},
  author={Loaiza, Ignacio and Khah, Alireza Marefat and Wiebe, Nathan and Izmaylov, Artur F},
  journal={Quantum Science and Technology},
  volume={8},
  number={3},
  pages={035019},
  year={2023},
  publisher={IOP Publishing}, 
  doi={https://doi.org/10.1088/2058-9565/acd577},
}

@misc{shukla2025practicalquantumphaseestimation,
      title={Towards Practical Quantum Phase Estimation: A Modular, Scalable, and Adaptive Approach}, 
      author={Alok Shukla and Prakash Vedula},
      year={2025},
      eprint={2507.22460},
      archivePrefix={arXiv},
      primaryClass={quant-ph},
      url={https://arxiv.org/abs/2507.22460}, 
}

@misc{blunt_monte_2025,
	title = {A {Monte} {Carlo} approach to bound {Trotter} error},
	url = {http://arxiv.org/abs/2510.11621},
	doi = {10.48550/arXiv.2510.11621},
	urldate = {2025-10-17},
	publisher = {arXiv},
	author = {Blunt, Nick S. and Ivanov, Aleksei V. and Bay-Smidt, Andreas Juul},
	month = oct,
	year = {2025},
	note = {arXiv:2510.11621 [quant-ph]},
}

@article{childs_theory_2021,
	title = {A {Theory} of {Trotter} {Error}},
	volume = {11},
	issn = {2160-3308},
	url = {http://arxiv.org/abs/1912.08854},
	doi = {10.1103/PhysRevX.11.011020},
	language = {en},
	number = {1},
	urldate = {2025-09-01},
	journal = {Physical Review X},
	author = {Childs, Andrew M. and Su, Yuan and Tran, Minh C. and Wiebe, Nathan and Zhu, Shuchen},
	month = feb,
	year = {2021},
	note = {arXiv:1912.08854 [quant-ph]},
	pages = {011020},
}

@article{suzuki_general_1991,
	title = {General theory of fractal path integrals with applications to many-body theories and statistical physics},
	volume = {32},
	issn = {0022-2488, 1089-7658},
	url = {https://pubs.aip.org/jmp/article/32/2/400/229229/General-theory-of-fractal-path-integrals-with},
	doi = {10.1063/1.529425},
	language = {en},
	number = {2},
	urldate = {2025-09-01},
	journal = {Journal of Mathematical Physics},
	author = {Suzuki, Masuo},
	month = feb,
	year = {1991},
	pages = {400--407},
}

@article{rajagopal_generalization_1999,
	title = {Generalization of the {Lie}-{Trotter} {Product} {Formula} for q-{Exponential} {Operators}},
	volume = {257},
	issn = {03759601},
	url = {http://arxiv.org/abs/cond-mat/9903106},
	doi = {10.1016/S0375-9601(99)00295-9},
	language = {en},
	number = {5-6},
	urldate = {2025-09-01},
	journal = {Physics Letters A},
	author = {Rajagopal, A. K. and Tsallis, Constantino},
	month = jul,
	year = {1999},
	note = {arXiv:cond-mat/9903106},
	pages = {283--287},
}

@misc{mehendale_estimating_2025,
	title = {Estimating {Trotter} {Approximation} {Errors} to {Optimize} {Hamiltonian} {Partitioning} for {Lower} {Eigenvalue} {Errors}},
	url = {http://arxiv.org/abs/2312.13282},
	doi = {10.48550/arXiv.2312.13282},
	urldate = {2025-09-04},
	publisher = {arXiv},
	author = {Mehendale, Shashank G. and Martínez-Martínez, Luis A. and Kamath, Prathami Divakar and Izmaylov, Artur F.},
	month = aug,
	year = {2025},
	note = {arXiv:2312.13282 [physics]},
}

\end{document}